\documentclass[aps,prb,twocolumn,superscriptaddress]{revtex4}

\usepackage{amsmath,amssymb}
\usepackage{graphicx}
\usepackage{epstopdf}
\usepackage{bm}
\usepackage{color}
\usepackage{mhchem}

\newcommand{\lsco}     {La$_{2-x}$Sr$_x$CuO$_4$}
\newcommand{\lbco}     {La$_{2-x}$Ba$_x$CuO$_4$}
\newcommand{\lesco}    {La$_{1.8-x}$Eu$_{0.2}$Sr$_x$CuO$_4$}
\newcommand{\lescoa}   {La$_{1.67}$Eu$_{0.2}$Sr$_{0.13}$CuO$_4$}
\newcommand{\lescob}   {La$_{1.6}$Eu$_{0.2}$Sr$_{0.2}$CuO$_4$}

\newcommand{\lmsco}    {La$_{2-x-y}$M$_y$Sr$_x$CuO$_4$}

\newcommand{\tlt}      {$T_\text{LT}$}
\newcommand{\tht}      {$T_\text{HT}$}
\newcommand{\la}       {$^{139}$La}
\newcommand{\laslr}    {$^{139}T_1^{-1}$}
\newcommand{\slr}      {$T_1^{-1}$}
\newcommand{\ltot}     {LTO $\rightarrow$ LTT}
\newcommand{\htot}     {HTT $\rightarrow$ LTO}

\begin{document}

\title{Structural transitions in a doped lanthanum cuprate}

\author{S.-H. Baek}
\affiliation{IFW-Dresden, Institute for Solid State Research,
PF 270116, 01171 Dresden, Germany}

\author{P. C. Hammel}
\affiliation{Department of physics, The Ohio State University,
Columbus, OH 43210, USA}

\author{M. H\"ucker}
\affiliation{Condensed Matter Physics and Materials Science Department, Brookhaven National Laboratory, Upton, New York 11973, USA}
\author{B. B\"uchner}
\affiliation{IFW-Dresden, Institute for Solid State Research,
PF 270116, 01171 Dresden, Germany}
\affiliation{Institut f\"ur Festk\"orperphysik, Technische Universit\"at
Dresden, 01062 Dresden, Germany}
\author{U. Ammerahl}
\affiliation{Laboratoire de Chimie des Solides, Universit\'{e} Paris-Sud,
91405 Orsay Cedex, France}
\author{A. Revcolevschi}
\affiliation{Laboratoire de Chimie des Solides, Universit\'{e} Paris-Sud,
91405 Orsay Cedex, France}
\author{B. J. Suh}
\email[Corresponding Author:~]{bjsuh@catholic.ac.kr}
\affiliation{Department of Physics, The Catholic University of
Korea, Bucheon 420-743, Korea} 
\date{\today}

\begin{abstract}
\la\ NMR and relaxation measurements have been performed
on \lesco\ ($x=0.13$ and 0.2) single crystals. The temperature
dependence of the \la\ NMR
spectra in all the structural phases [high-temperature tetragonal (HTT) 
$\rightarrow$ low-temperature orthorhombic (LTO) $\rightarrow$ low-temperature 
tetragonal (LTT)] 
reveals the non-vanishing tilt angle of the CuO$_6$ octahedra in
the HTT phase, opposed to the case of \lsco\ where the tilt angle
disappears immediately above the transition. Since \la\ relaxation data provide
evidence of the thermodynamic critical fluctuations associated with the structural phase
transitions, \htot\ and \ltot, we conclude that the structural transitions in
Eu-doped \lsco\ should be of the order-disorder type rather than of the
displacive type observed in \lsco. The change of the nature of the structural transitions
caused by doping with Eu appears to be consistent with the \ltot\ transition that is
absent in \lsco.
\end{abstract}


\maketitle

\section{Introduction}

The intrinsic structural instability and
its strong influence on superconductivity observed in
La$_{2-x}$(Ba,Sr)$_x$CuO$_4$ and in rare-earth-doped \lmsco\ (M = Nd,
Eu) compounds\cite{axe89,buchner94} emphasize the
crucial role of local structure in high-temperature superconductivity.  
Moreover, the effects of structural distortions in these materials have been manifested 
in the formation of static stripe order.\cite{vojta09,hucker12} 
Despite a great deal of work, however, some issues remain controversial
and the detailed nature of the local structure of lanthanum cuprates is not yet fully
understood.

As is well known from x-ray and neutron diffraction studies, for hole
concentration $x$ above a certain value in \lbco\ and \lmsco,
a sequence of structural phase transitions (SPTs) occurs on lowering
temperature ($T$) : high-temperature tetragonal (HTT) phase $\rightarrow$
low-temperature orthorhombic (LTO) phase $\rightarrow$ low-temperature
tetragonal (LTT) phase.
While there is a consensus that these rich structural phases are associated with the
subtle changes in the tilt angle and/or tilt axis of the CuO$_6$
octahedra,
there has been much debate on whether or not the macroscopic structure
corresponds to the local one.
In the average structure
model obtained by diffraction studies, the \htot\ is characterized
mainly by the tilt angle of the CuO$_6$ octahedra, whereby there is no
tilt of the CuO$_6$, i.e., the CuO$_2$ planes are flat, in the HTT phase.
In addition, the \ltot\ transition is ascribed to the rotation of the
tilt axis of CuO$_6$ octahedra through 45$^\circ$
in the average structure model,\cite{axe94,friedrich96,simovic03a,hucker04} in which
the change of the direction of the tilt
axis seems to have a profound effect on the electronic
properties, resulting in the stabilization of
static stripe order and the destruction of
superconductivity.\cite{buchner94,klauss00,vojta09}

The other view is called the local structure model, in which the direction of the
local tilts does not change at the \ltot\ transition and the
LTO structure results from coherent superposition of the local LTT
variants.\cite{billinge94,haskel00,han02}
In this model, the local tilts persist even in the HTT phase \cite{wakimoto06} 
but the tilt axes are in 
disorder resulting in the average HTT phase.
A similar argument can be found in
Ref.~\onlinecite{haskel96} although a discrepancy
exists between the conclusions in Refs. \onlinecite{billinge94} and
\onlinecite{haskel96} with respect to the direction of the local CuO$_6$ tilt.

The main discrepancy between the average and local structure model
lies in whether the tilt of the CuO$_6$ exists or not in the HTT phase.
Thus the key to discriminate between two models is to measure
the local tilt angle in the HTT phase.
\la\ nuclear magnetic resonance (NMR) is a well-known local probe which is
extremely sensitive to certain details of the local structure of lanthanum
cuprates.
In fact, the non-zero angle ($\alpha$) in the LTO and LTT phases between the
principal axis of the
electric field gradient (EFG) at the La site and the crystal $c$-axis
has been well known and is directly induced
from the tilt angle ($\phi$) of the CuO$_6$ octahedra.
Here we report the \la\ NMR studies of \lesco\ single
crystals, providing evidence for the non-vanishing tilt angle of the CuO$_6$
octahedra in the HTT phase supporting the local structure model.
Furthermore, distinct behaviors of the \la\ spin-lattice relaxation rate at the \htot\
and \ltot\ transitions were observed, consistent with the fact that the 
former is a second 
order and the latter a first order.\cite{axe94,cherny95}

\section{Sample preparation and experimental details}

The single crystals were grown using the traveling solvent floating
zone (TSFZ) method under an oxygen pressure of 3 bar.
From x-ray diffraction and thermal expansion studies, we observed
the \ltot\ transition at $T_\text{LT}= 134(2)$ and $110(10)$ K for $x=
0.13$ and 0.20, respectively, and the \htot\ transition at $T_\text{HT}=
225(5)$ K for $x = 0.20$.

\la\ NMR and spin-lattice relaxation measurements
were performed in \lesco\
single crystals with $x = 0.13$ and 0.20.
\la\ (nuclear spin $I = 7/2$) NMR spectra were obtained by sweeping external magnetic
field, $H$, at fixed resonance frequencies, $\nu_0$ in the temperature range
4.2--360 K.
\la\ spin-lattice relaxation rates were measured
at the central transition ($+1/2 \leftrightarrow -1/2$) by monitoring
the recovery of magnetization after saturation with a single $\pi/2$ pulse.
Since a common recovery law cannot explain the data for the whole
temperature range investigated,
we obtained effective spin-lattice relaxation rates \laslr\ by fitting the
recovery data for the first decade in the whole temperature range
investigated to the stretched exponential function:
$[M(\infty)-M(t)]/M(\infty) = a\exp[-(t/T_1)^{0.5}]$ where $M$ is the nuclear
magnetization and $a$ a fitting parameter.

\section{Results and Discussion}

Figure~1 shows the \la\ spectra obtained in \lescoa\ at 29 MHz in the LTT phase
(4 K), for $H$ applied parallel
and perpendicular to the $c$ axis. These spectra represent typical quadrupole-perturbed
ones which are observed in a single crystal with $I = 7/2$.
Note that the flat background between satellites for
$H \parallel c$ indicates the high quality of the single crystal.
\begin{figure}
\centering
\includegraphics[width=\linewidth]{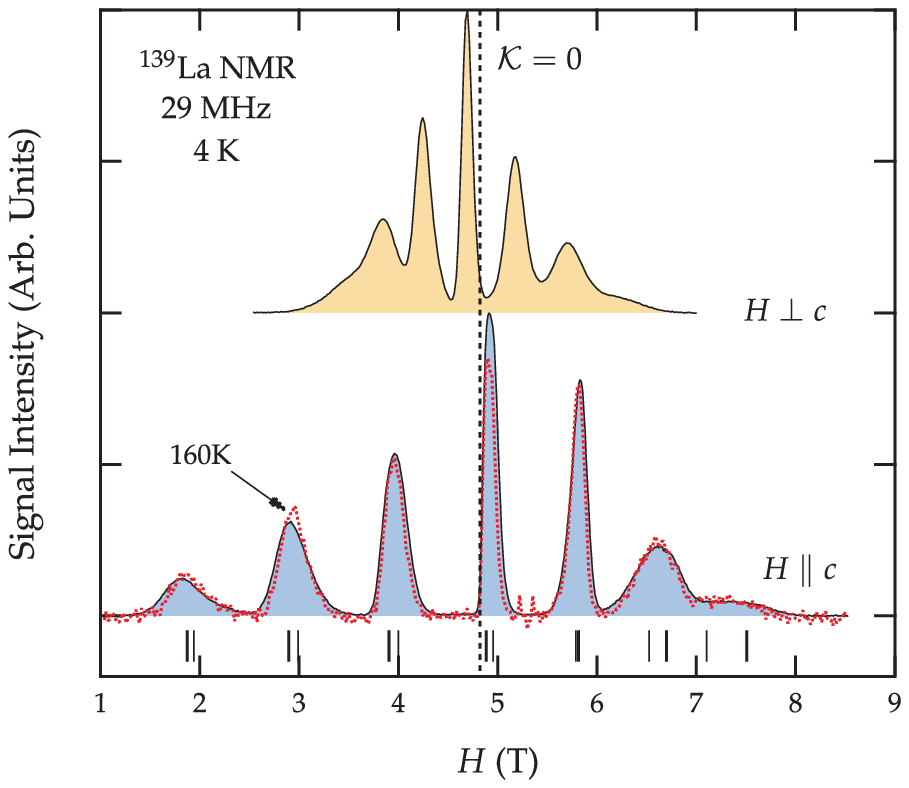}
\caption{\label{fig:spec} \la\ NMR spectra in a \lescoa\ single
crystal measured at $\nu_0 = 29$~MHz and 4 K (LTT phase)
for two different orientations of the applied field ({\bf H})
with respect to the crystal $c$-axis. Vertical dotted line indicates the zero 
Knight shift position.  
The lines under the spectra are the marks of the positions of the
satellites calculated for $\alpha = 11^{\circ}$ (thick line)  and
$17^{\circ}$ (thin line)  by assuming $\nu_Q = 6$~MHz and $\eta = 0$.
Spectrum measured in the LTO phase (160 K) for $H \parallel c$ (dotted line) 
shows a similar
asymmetrical pattern with that at 4 K, indicating that the distribution of the
local tilt angle $\alpha$ is only slightly reduced above the \ltot\ transition.
}
\end{figure}
In addition to the satellites which result from the first order quadrupole
effect, there is also a second order quadrupole effect which shifts the central
transition ($1/2\leftrightarrow -1/2$) depending on the tilt angle $\alpha$
between the principal
axis of the electric field gradient (EFG) (i.e., the axis of $V_{zz}$) and the
external field $H$:
\begin{equation}
\nu_0 = \gamma_N (1 + \mathcal{K})H + \frac{15\nu_Q^2}{16\nu_0}(1-\mu^2)(1-9\mu^2),
\end{equation}
where $\gamma_N = 6.014$ MHz/T is the nuclear gyromagnetic ratio
of \la, $\mathcal{K}$ the Knight shift, $\nu_Q$ the quadrupole frequency, and
$\mu \equiv \cos \alpha$.
Since the second order shift should
vanish when $\alpha=0$ and $\mathcal{K}$ is very small for the \la\ in
La-based cuprates,\cite{baek12a} the considerable shift of the central line for
$H\parallel c$ from the
resonance field for $\mathcal{K}=0$
indicates that $V_{zz}$ is tilted out of the $c$ axis by the
angle $\alpha$.
Also, we find that the spectra display asymmetric features in
the position and the linewidth
of mirror satellites ($m \leftrightarrow m-1/2$ and $-m \leftrightarrow
-m+1/2$) with respect to the central transition. This implies that the tilt angle $\alpha$
is considerably distributed. To support this, we calculated the resonant fields of
the transitions for $\alpha=11^\circ$ and $17^\circ$ by assuming $\nu_Q=6$ MHz
and the asymmetry parameter $\eta=0$. The results are denoted in Fig. 1 as the
thick ($11^\circ$) and thin ($17^\circ$) lines under the
spectra, which accounts for the asymmetric satellite spectra very well.
Interestingly, the spectra measured at 160 K in the LTO phase are almost
identical with those at 4 K, with slightly reduced linewidths.  This indicates
that the distribution of the tilt angle is
robust in a wide range of temperatures, even above the \ltot\ transition.

\begin{figure}
\centering
\includegraphics[width=\linewidth]{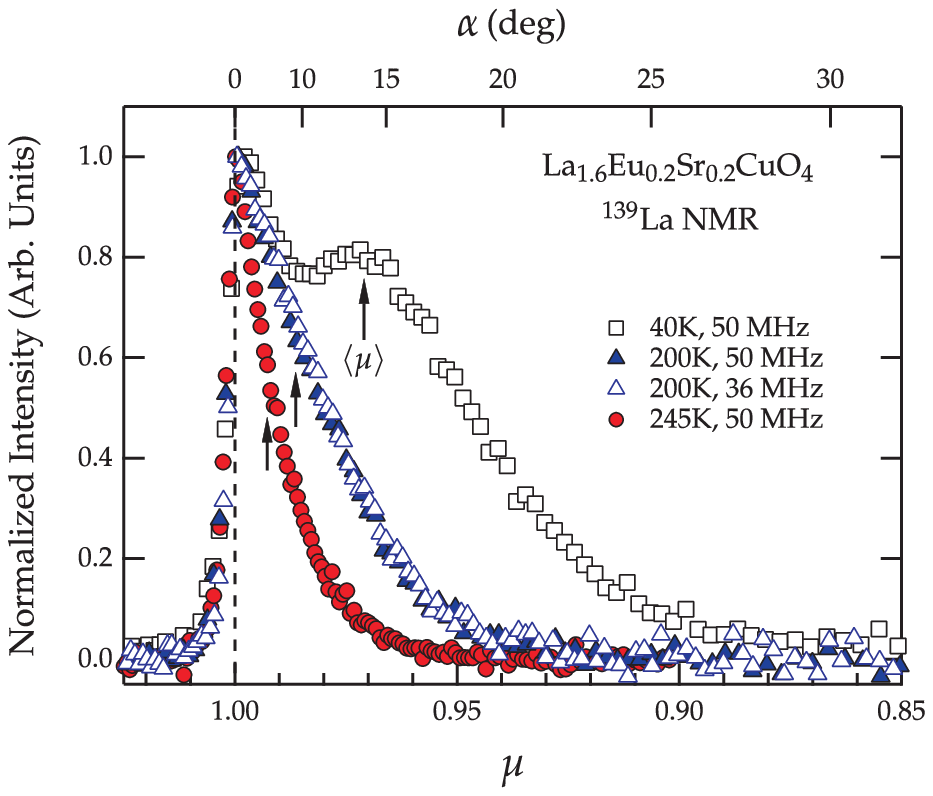}
\caption{\label{fig:mu}Temperature dependence of \la\ NMR spectrum at the central transition
for \lescob. The external field $H$ has
been transformed to $\mu \equiv \cos \alpha$ using the Eq.~(1).  The
arrows are marks for the average value $\langle\mu\rangle$, 
which was obtained from the center of gravity for each spectrum.
Note that $\langle \mu\rangle$ is still finite at 245 K $> T_\text{HT}$.}

\end{figure}

Since $\alpha$ can be obtained from the second order
shift of the central line according to Eq.~(1), from now on we focus on the $T$-dependence
of the central line in order to get detailed information of the tilt angle of the
CuO$_6$ octahedra.
In Fig.~2, we show the evolution of the central NMR line on
lowering $T$ in \lescob. Here the field $H$ was converted to $\mu=\cos\alpha$
using Eq. (1) by assuming $\mathcal{K}=0$. Although there should be a distribution of 
$\nu_Q$ which may cause the extension of the spectrum to $\mu>1$ in Fig. 2, 
the width ($\sim 1$ MHz) of the \la\ NQR spectrum\cite{suh99} should give the 
symmetric broadening of the NMR central line of $\sim20$ kHz at $\nu_0=50$ MHz 
for a given angle $\alpha$
according to Eq.~(1). This allows us to treat $\nu_Q$ as a constant for practical 
purposes.  
Note that two spectra for different frequencies measured at 200 K coincide.
This indicates that the shift of the central line is mostly of a quadrupolar origin rather
than of a magnetic one.
At 40 K, we find that the line is clearly resolved to two.
The magnetic shift ($\mathcal{K}$), \slr, and $T_2^{-1}$ for the two sites are found
to be identical, indicating that the two lines represent local structural
inhomogeneity rather than different magnetic properties.
Indeed, two distinct lines are attributed to the
so called buckling of the CuO$_2$ plane which results from the
tilt of the CuO$_6$ octahedra along [010]$_\text{LTO}$ whose direction
alternates by $180^\circ$ along $[110]_\text{LTO}$, which was recently demonstrated in
\lsco (LSCO).\cite{baek12a}

For quantitative analysis, we rewrite Eq. (1) in the form
\begin{gather}
\frac{\delta H}{H} \cong \mathcal{K} + \frac{C}{\nu_0^2} \\
\intertext{with}
C = \frac{15\nu_Q^2}{16}(1-\mu^2)(1-9\mu^2).
\end{gather}
In writing Eq.~(2), we used $\nu_0 \cong \gamma_N H$ and
$\delta H \equiv \nu_0/(\gamma_N) - H$.  Thus, by measuring spectra at
different frequencies, one can
separate the shift into the contribution of the magnetic ($\mathcal{K}$)
and the quadrupole origin.
The inset of Fig.~3 demonstrates how the magnetic shift ($\mathcal{K}$) and the
quadrupolar contribution ($C$) can be separated using Eq.~(2) and the
actual dominance of the quadrupole contribution with negligibly small
$\mathcal{K}$. Therefore, the
average shift of the central line is indeed equivalent to a measure of the
average tilt angle $\langle\alpha\rangle$.

\begin{figure}
\centering
\includegraphics[width=\linewidth]{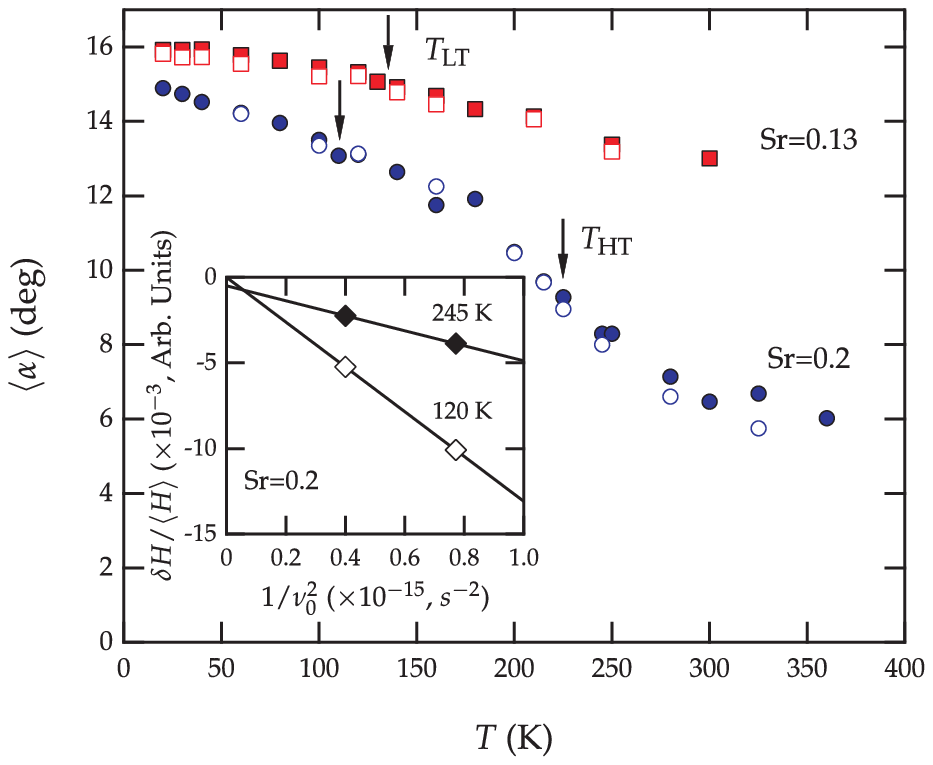}
\caption{Temperature dependence of the average angle $\langle\alpha\rangle$.
$\langle\alpha\rangle$ does not show an anomaly at both $T_\text{LT}$ and
$T_\text{HT}$ (indicated by arrows), persisting even at temperatures that are much higher than
$T_\text{HT}$ for Sr=0.2. Inset shows that the shift of spectra is mostly
of quadrupolar origin with the small Knight shift ($C/\nu_0^2=0$) [see Eq. (2)].}
\label{alpha}
\end{figure}

We plot $\langle\alpha\rangle$ converted from the average field
$\langle H\rangle$ for the two doped samples.  In Fig.~3, we actually plot two sets of
$\langle\alpha\rangle$:  the one (open symbols) is obtained from the values of
$\langle H\rangle$ of spectra measured at two
different frequencies using Eq.~(2) as demonstrated in the inset, and the
other (closed symbols) is obtained simply from $\langle H\rangle$ of a single spectrum
measured at one frequency by assuming $\mathcal{K} = 0$ in Eq.~(2).
No noticeable difference between the open and closed symbols
indicates that $\mathcal{K}$ is negligibly small.

The main feature in Fig.~3 is non-vanishing $\alpha$ at high
$T$ in the HTT phase.  This is clearly contrasted with the picture described
by the average structure model, but
rather supports the local structure model.
One may claim that the non-zero $\alpha$ is attributable to the
experimental error in aligning the $c$-axis of the sample along
the direction of the applied field $H$.
However, the misalignment of the sample should give rise to a
narrowing of the line at high $T$ at both edges of the spectrum,
i.e., the line should shrink with respect to the center corresponding
to the misalignment angle.
Figure 2 clearly shows that the left edge of the spectrum corresponding to the
smaller $\alpha$ does not change while a significant narrowing of the
spectrum takes place with increasing $T$ at the right edge of the spectrum corresponding
to the larger $\alpha$.
Thus, we conclude that the non-zero $\alpha$ is totally irrelevant to the experimental
error.

\begin{figure}
\centering
\includegraphics[width=\linewidth]{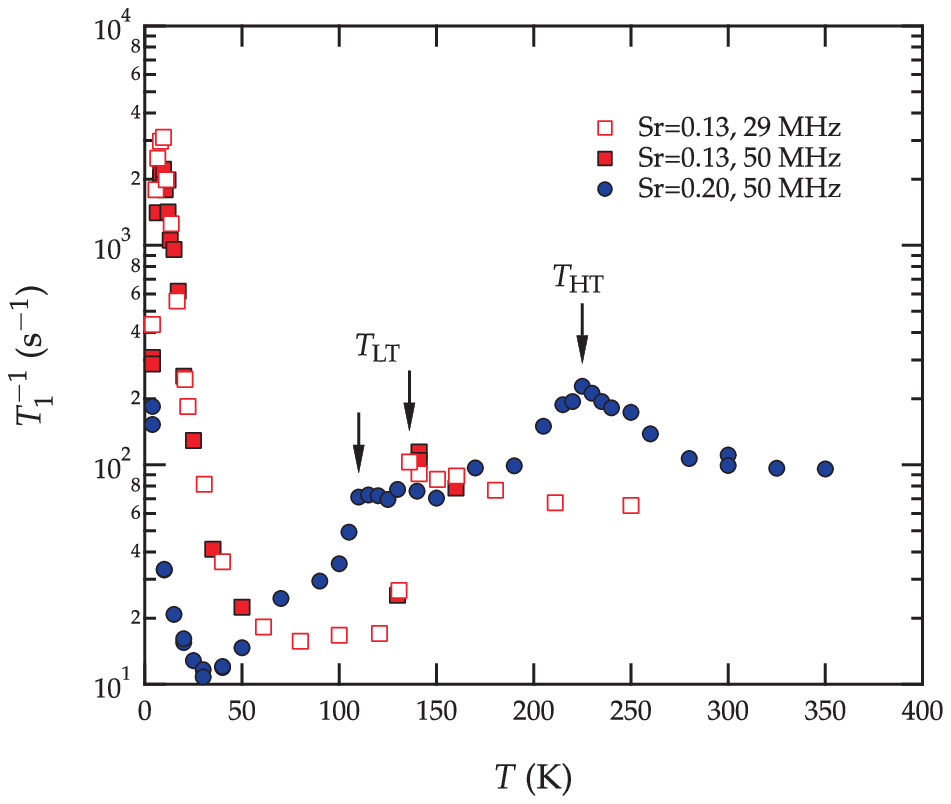}
\caption{\la\ spin-lattice relaxation
rate \slr\ as a function of temperature measured at the central transition
for $H \parallel c$ in \lesco\ single crystals. Data for Sr=0.13 are identical
with those reported in Ref.~\onlinecite{suh00}. While the transition at
\tlt\ causes
the sharp drop of \laslr\ indicating first order transition, the \slr\ anomaly at
\tht\ is clearly continuous, being consistent with second order transition.}
\label{slr}
\end{figure}

A remarkable finding is that there is no visible anomaly of $\langle\alpha\rangle$ at
both \tlt\ and \tht\ in Fig. 3.
In contrast, we find that the collective modes
associated with the phase
transitions are clearly picked up by the \la\ spin-lattice relaxation rate
\slr\ as shown in Fig. 4. Namely, \slr\ displays distinct
features at the structural transitions: a sudden decrease at \tlt\ and a relatively
sharp anomaly at \tht\ above which a clear upturn of \slr\ precedes.
A strong enhancement observed at low temperatures below $\sim 50$ K for both samples
is attributed to glassy
spin freezing phenomenon. Note that for Sr=0.2 the low-$T$ enhancement of
\slr\ is significantly suppressed by an order of magnitude.

The sudden drop of \slr\ at \tlt\ with essentially no
enhancement above is in contrast to the continuous and relatively sharp peak
of \slr\ at \tht.  Since the NMR spin-lattice relaxation rate reflects the critical behavior of
collective modes at a phase transition, 
the contrasting behaviors of \slr\ at \tlt\ and \tht\ suggest a different
thermodynamic nature of the structural transitions.   This is in
good agreement with
the observation that the \ltot\ transition in the heavily hole-doped sample
is discontinuous (first order) where the
well-defined critical behavior of the order parameter does not exist, while the
\htot\ is dominantly a second order \cite{buechner93,axe94,suh99} although weakly first
order nature may be present, as observed in the underdoped LSCO sample.\cite{baek12a}


Therefore, the absence of an anomalous change of $\langle\alpha\rangle$ at \tht\ and \tlt\
cannot be due to an
inhomogeneous structural mixture or simple tilt angle disorder caused by
dopants, but should reflect characteristics of the structural phase transitions.
For the \ltot\ transition, no change of $\langle\alpha\rangle$ at
\tlt\ is actually expected for both the average and local structure models,
since the difference of the two models is the direction of the local tilt axis
in the LTT and LTO phases, not the average tilt angle itself.

On the other hand, for the \htot\ transition, the significant tilt
angle which remains in the HTT phase suggests that the \htot\ is
characterized dominantly by an order-disorder type transition where the HTT
phase results from disordered LTO structures.
This contrasts sharply with the displacive transition observed in LSCO in
which the CuO$_2$ plane becomes flat in the HTT phase.\cite{braden01,baek12a}
This implies that doping Eu in LSCO modifies the nature
of the \htot\ transition. Since the LTT
phase at low $T$ does not occur in LSCO, it is possible that the different type
of the \htot\ transition is relevant to the occurrence of the LTT phase.
It was argued that
the thermal conductivity peak that appears in La$_2$CuO$_4$ and
non-superconducting rare earth-doped LSCO is suppressed in
superconducting LSCO due to enhanced scattering of
the heat carrying phonons with soft phonons.\cite{baberski98,hess03}
We find that a similar argument could be made for the different nature
of the \htot\ caused by Eu doping.
The dominant soft phonon modes in LSCO are actually consistent with the
second order displacive \htot\ structural transition observed in
LSCO,\cite{boni88,baek12a} considering that displacive materials are characterized by low
anharmonicity or quasi-harmonicity.\cite{rehwald73,cowley80,nakanishi82}
By doping Eu, however, strong anharmonicity of
the vibrations is likely introduced so that the soft mode is overdamped, being
insufficient to drive a displacive transition and resulting in an
order-disorder type transition instead.
In this case, however, it would be quite natural that the transition is
imperfect due to a structural inhomogeneity caused by dopants.
Indeed, the fact that $\langle\alpha\rangle$
decreases continuously toward zero with increasing $T$ as shown in Fig. 3
may suggest that \textit{short-range precursor order}\cite{bruce80} with a displacive nature
occurs at $T\gg T_\text{HT}$, yielding to the long-range order-disorder
transition at \tht.


\section{Conclusion}

We studied \la\ NMR of \lesco\ ($x=0.13$ and 0.2) as a function
of temperature. The $T$-dependence of \slr\ confirms the critical modes
associated with the structural transitions, showing first order \ltot\ and second
order \htot\ transitions.  An important finding is that the local tilt angle of the CuO$_6$
octahedra does not vanish in
the HTT phase, which is consistent with the local structure model. This is opposed to
\lsco\ in which the average structure model is strongly supported. This
indicates that Eu-doping causes the change of the nature of the structural
transitions from a
displacive \htot\ transition in LSCO to an order-disorder one in Eu-doped LSCO.
This unusual sensitivity of the lattice dynamics in La cuprates is
attributed to the soft phonon modes that are highly susceptible to dopants.

\section*{Acknowledgement}

This work was supported by the Basic Science Research Program through
the National Research Foundation of Korea (NRF) funded by the
Ministry of Education, Science and Technology
(NRF-2008-314-C00123). This work was supported in part by the US
DOE. One of authors (B.J.S.) acknowledges the support from the
2008 Research Fund of the Catholic University of Korea.
M.H. acknowledges support by the Office of Science, U.S. Department of Energy
under Contract No.~DE-AC02-98CH10886.
\bibliography{mybib}

\end{document}